\documentstyle[12pt]{article}

\newcommand{\lsim}{\mathrel{\lower4pt\hbox{$\sim$}}
\hskip-12.5pt\raise1.6pt\hbox{$<$}\;}

\newcommand{\gsim}{\mathrel{\lower4pt\hbox{$\sim$}}
\hskip-12.5pt\raise1.6pt\hbox{$>$}\;}

\begin{document}
 
%=======================================================================
%   BEGIN TITLEPAGE
%=======================================================================
\begin{titlepage}

\baselineskip=18pt 

\begin{flushright}
{\bf UR-1384} \\
{\bf MADPH-96-968}
\end{flushright}

\bigskip

\begin{center}

{\large\bf Parity Violation in Bottom Quark Pair Production \\
at Polarized Hadron Colliders}

\vspace{0.36in}

Chung Kao$^{a,b}$, David Atwood$^c$ and Amarjit Soni$^d$ 

\bigskip

$^a${\sl Department of Physics and Astronomy, University of Rochester, \\
Rochester, NY 14627, USA} \\
$^b${\sl Department of Physics, University of Wisconsin, 
Madison, WI 53706, USA} \\
$^c${\sl Theory Group, CEBAF, Newport News, VA 23606, USA}\\
$^d${\sl Department of Physics, Brookhaven National Laboratory, 
Upton, NY 11973, USA}

\end{center}

\vspace{0.36in}

\begin{abstract}

Parity violation induced by 
the chromo-anapole form factor of the bottom quark, 
generated from weak corrections, 
is studied in polarized hadron collisions.
The forward-backward asymmetry in the bottom quark pair production 
at polarized $pp$ and $p\bar{p}$ colliders is evaluated 
in the Standard Model and in a Two Higgs Doublet Model 
to examine the effects of parity violation.
In the models studied, 
promising results are found for polarized $p\bar{p}$ colliders.

\end{abstract}

\end{titlepage}
%

%=======================================================================
%   MAIN TEXT
%=======================================================================
\baselineskip=24pt % Double Space

%-----------------------------------------------------------------------
%   1. Introduction
%-----------------------------------------------------------------------
\newpage

\section{Introduction}

The large production cross section of bottom quark pairs ($b\bar{b}$) 
at polarized hadron colliders \cite{Contoug,Karliner} 
coupled with the relatively long lifetime of the $b$-quark, 
implies that detailed experimental study of the properties 
of the $b$-quark at those facilities could also be very useful. 
Since the CKM mixing angle $V_{tb}$ is close to one, 
the $b$ quark couples rather readily to the virtual top quark 
in loop diagrams.
Now the top quark is so heavy ($m_t\sim175$ GeV) \cite{CDF,D0} that 
the $b$-quark becomes very sensitive to electroweak radiative corrections 
as these corrections often tend to grow with the 
virtual quark mass \cite{HWS}.
Precision studies of the $b$'s are thus very useful 
in testing the Standard Model (SM) and in searching for new physics.

In the hadronic environment, 
while the production cross sections are high, 
a quantitative understanding of the effects of QCD 
can be a very difficult challenge. For this reason, 
in testing the SM and in searching for clues of new physics, 
it is perhaps better to focus on observables that tend to be robust 
to QCD corrections. For that reason, 
in general, the production cross section is not a good observable. 
We propose to focus instead on signatures of parity violation 
in $b$-quark production 
since QCD corrections cannot generate parity violation.
Parity violating asymmetries that are ratios of cross-sections
should be less sensitive to QCD corrections as well as
to the uncertainties in the parton distribution
functions. 
Furthermore, in study of parity violation, one may be able to make use 
of polarized incident $p(\bar{p})$ beams \cite{Bunce}. 
We will concentrate on one type of parity violating observable,  
that is the forward-backward asymmetry. 
In its differential form, it is defined as:
\begin{equation}
\delta {\cal A}(M_{b\bar{b}}) \equiv 
\frac{ d\sigma_F/dM_{b\bar{b}} -d\sigma_B/dM_{b\bar{b}} }
     { d\sigma_F/dM_{b\bar{b}} +d\sigma_B/dM_{b\bar{b}} } 
\label{eq:deltaA}
\end{equation}
where the subscripts $F$ and $B$ stand for the forward 
and the backward hemispheres respectively.

For the reactions of interest to us, 
i.e. $pp\to b\bar{b} +X$ or $p\bar{p} \to b\bar{b}+X$, 
we consider 
two sources of contributions 
to such a parity violating observable. These are 
the chromo-anapole form factor of the $gb\bar{b}$ vertex 
(note $g \equiv gluon$)
and the tree level electroweak process $q\bar{q} \to Z \to b\bar{b}$. 
The latter contribution ($Z$ exchange) is significant 
only when the invariant mass ($M_{b\bar{b}}$) of the $b\bar{b}$ pair 
is close to the $Z$ mass ($M_Z$). 
It can be removed, if necessary, by imposing an
appropriate cut on the $M_{b\bar{b}}$.
Thus, the chromo-anapole form factor is a very important contributor 
to the parity violating signal and consequently it is our primary focus.

% ------------ New ------------

Recently, several suggestions have been made to study parity violation 
asymmetries in polarized hadron collisions for
(1) the production of one jet, two jets, and two jet plus photon 
in the SM \cite{Cheng};  
(2) the production of $W^\pm$ and $Z$ in the SM \cite{Bourrely&Soffer}; and 
(3) the inclusive production of one jet with a new handed interaction 
between subconstituents of quarks \cite{Taxil&Virey}.
These works primarily deal with interference of electroweak
and strong interactions on light quarks. 

In this Letter, we present the first study on parity violation
generated from one loop weak corrections to bottom quark pair production 
at polarized hadron colliders. The key difference with the works
in Refs. \cite{Cheng,Bourrely&Soffer,Taxil&Virey} is, 
as alluded to in the opening paragraph, 
that the $b\bar{b}$ pair in the final state in our study 
is a very sensitive tool of the effects of the top quark 
which in turn is sensitive to non-standard effects.  
Specifically, we will evaluate 
the forward-backward asymmetry in the bottom quark pair production 
at polarized $pp$ and $p\bar{p}$ colliders 
in the Standard Model and in a Two Higgs Doublet Model 
to examine the effects of parity violation 
from the interactions of $b\bar{b}$ with spin-1 and spin-0 fields.

%-----------------------------------------------------------------------
%   2. Form Factors
%-----------------------------------------------------------------------

\section{Form Factors}

Let us write the $gb\bar{b}$ vertex as
\begin{equation}
-ig_s \bar u(p_1) T^a \Gamma^\mu v(p_2)
\end{equation}
where $g_s={}$the strong coupling, $T^a={}$the SU(3)
matrices, $u(p_1)$ and $v(p_2)$ are the Dirac spinors of $b$ and $\bar{b}$
with outgoing momenta $p_1$ and $p_2$ and $k=p_1+p_2$ is the momentum
of the gluon. At the tree level $\Gamma^\mu_0=\gamma^\mu$. The 1-loop
vertex function can be expressed as
\begin{eqnarray}
\Gamma^\mu 
& = & \gamma^\mu [A(k^2)-B(k^2) \gamma_5] \nonumber \\
& & + (p_1-p_2)^\mu [C(k^2) - D(k^2) \gamma_5] \nonumber \\
& & + (p_1+p_2)^\mu [E(k^2) - F(k^2) \gamma_5]  
\label{eq:FF1}
\end{eqnarray}
Current conservation demands that $E = 0$ and $B = -k^2F(k^2)/2m_b$, 
where $k=p_1+p_2$ and $k^2=M^2_{b\bar{b}}=\hat s$ 
in the $b\bar{b}$ center of mass (CM) frame. 
Applying the Gordon identities we can re-write 
\begin{eqnarray}
\Gamma_\mu 
& = & F_1(k^2) \gamma^\mu -F_2(k^2) i\sigma^{\mu\nu} k_\nu \nonumber \\
&   & +a(k^2)  \gamma_\nu \gamma_5 ( k^2 g^{\mu\nu} -k^\mu k^\nu)
      +d(k^2)  i\sigma^{\mu\nu} k_\nu \gamma_5
\label{eq:FF2}
\end{eqnarray}
where $F_1(k^2) = A(k^2) + 2m_b C(k^2)$, $F_1(0)={}$the
chromo-charge; $F_2(k^2)=C(k^2)$, $F_2(0) ={}$the anomalous
chromo-magnetic moment; $a(k^2) = -B(k^2)/k^2 = F/(2m_b)$,
$a(0)={}$chromo-anapole moment; $d(k^2)=D(k^2)$, and $d(0)={}$the
chromo-electric dipole moment. 

In the SM, the dominant one loop weak corrections that contribute to 
the chromo-anapole moment arise from diagrams with the $W^+$ 
and its Goldstone counterpart, the $G^+$. 
In many extensions of the SM, $e.\;g.$ in models
which contain more than one doublets of Higgs, 
there are charged Higgs bosons ($H^\pm$) contributing 
to the chromo-anapole form factor of fermions at the one-loop order. 
As is well known the simplest of such extensions consists of
two Higgs doublets \cite{Georgi} $\phi_1$ and $\phi_2$ 
with vacuum expectation values (VEVs) $v_1$ and $v_2$.
After symmetry breaking, 
there remain five physical Higgs bosons \cite{Guide}: 
a pair of singly charged Higgs bosons $H^{\pm}$, 
two neutral CP-even scalars $H$ (heavier) and $h$ (lighter), 
and a neutral CP-odd pseudoscalar $A$. 
The ratio of the two VEVs is usually expressed as 
$\tan\beta \equiv v_2/v_1$.

In our analysis, 
we have considered such a Two Higgs Doublet Model (THDM) 
with the Yukawa interactions of model II \cite{Model2}, 
which is required in the minimal supersymmetric model (MSSM)\footnote{
In the MSSM, though,there are additional contributions to 
the chromo-anapole moment from loop diagrams 
with the charginos and the squarks as well.}. \cite{Guide} 
In this model, 
one doublet ($\phi_1$) couples to down-type quarks and charged leptons 
while the other ($\phi_2$) couples to up-type quarks and neutrinos.
The one loop weak corrections from diagrams with the $W^+$, 
the $G^+$ and the $H^+$, 
yield dominant contribution to the b quark anapole moment from the
virtual top quark in the loop. 
In our calculations of these corrections we will set $V_{tb} = 1$ 
and also we will ignore CP violating effects.
We have employed the 't~Hooft--Feynman gauge for loop calculations
with $M_{G^\pm} = M_W$.

The forward-backward asymmetry in $b\bar{b}$ production 
is primarily driven by ${\rm Re} B(k^2)$ 
which is related to the chromo-anapole form factor $(a(k^2))$ 
via $B(k^2)=-k^2a(k^2)$. 
Therefore, in Fig.~1 we present ${\rm Re} B(k^2)$ 
coming separately from diagrams with the $G^\pm (B_G)$, 
the $W^\pm (B_W)$ and the $Z$ as well as the total contribution in the SM. 
As an illustration of a beyond the SM effect, 
we also show ${\rm Re} B(k^2)$ 
for THDM for $M_{H^+}=200$ GeV and various values of $\tan\beta$. 
The $Z$ boson exchange contribution to the $b$-chromo-anapole moment 
is very small since the coupling product $2g^b_V g^b_A$ is very small. 
The form factor ${\rm Re} B(k^2)$ is enhanced near $k^2 \sim 4m^2_t$,  
where $k^2 = \hat{s} = M^2_{b\bar{b}}$, 
due to the $t\bar{t}$ threshold which appears 
in diagrams with charged boson ($i.$ $e.$ $W^\pm$ or $H^\pm$) exchanges. 
Since $B(k^2)= -k^2a(k^2) = -\hat{s} a(\hat s)$, 
its value is expected to grow with $M_{b\bar{b}}$.
From Fig.~1 we see that numerically this increase sets in
for $M_{b\bar{b}} \gsim 500$ GeV.

In the THDM, the form factor $B(k^2)$ gets an additional contribution,
$B_H(k^2)$, from the charged Higgs boson. 
Clearly $B_H$ is a function of $M_{H^+}$ and $\tan \beta$, 
being proportional to $\cot^2\beta\{1-[(m_b/m_t)\tan^2\beta]^2\}$. 
Therefore, it has the same sign as $B_G$ 
for $\tan\beta<\sqrt{m_t/m_b}\sim 6$, 
but has an opposite sign for $\tan \beta>\sqrt{m_t/m_b}$.
Thus the effects of parity violation are enhanced in the THDM
if $\tan \beta$ is less than $\sqrt{m_t/m_b}$.
For $\tan\beta = \sqrt{m_t/m_b}$, the $B_H$ vanishes 
and the total $B(k^2)$ becomes that of the SM.

%-----------------------------------------------------------------------
%   3. Forward Backward Asymmetry in Hadron Collisions
%-----------------------------------------------------------------------

\section{Forward Backward Asymmetry}

Let us define the cross section for the sub-process, 
$q\bar{q} \to b\bar{b}$, in each helicity state of quarks 
in the initial state as 
\begin{equation}
\hat{\sigma}_{\lambda_1\lambda_2} \equiv 
\hat{\sigma}(q_{\lambda_1}\bar{q}_{\lambda_2}\to b\bar{b})
\end{equation}
where $\lambda_{1,2}$ represents a right-handed ($R$) or a
left-handed ($L$) helicity of the quark and the antiquark. 
The cross sections in the forward $(0\le\theta\le\pi)$ 
and the backward $(-\pi\le\theta\le0)$ directions 
are defined as
\begin{eqnarray}
\hat\sigma^F_{\lambda_1\lambda_2} & \equiv & 
\int^1_0 \frac{d\hat \sigma_{\lambda_1 \lambda_2}}{dz}dz \nonumber \\
\hat\sigma^B_{\lambda_1\lambda_2} & \equiv & 
\int^0_{-1} \frac{d\hat\sigma_{\lambda_1 \lambda_2}}{dz}dz \nonumber \\
\hat\sigma_{\lambda_1\lambda_2} & = & 
\hat\sigma^F_{\lambda_1\lambda_2} +\hat\sigma^B_{\lambda_1\lambda_2} 
\nonumber\\
\Delta\hat\sigma_{\lambda_1\lambda_2} & = & 
\hat\sigma^F_{\lambda_1\lambda_2} -\hat\sigma^B_{\lambda_1\lambda_2} 
\end{eqnarray}
where $z=\cos\theta$, with $\theta$ being the scattering
angle of the $b$ in the $b\bar{b}$ center of mass frame, 
$\hat\sigma$ is the total cross section of the $q\bar{q}$ subprocess 
and $\Delta\hat\sigma$ is the difference of the cross sections 
in the forward and the backward directions of the subprocess.
(Note that variables in the CM frame of the $q\bar{q}$ subprocess 
are denoted with a $\hat{\phantom{a}}$ on top of them).  

The tree level expressions for these cross sections  are  
\begin{eqnarray} 
\hat{\sigma}^0_{RL}
& = &\hat{\sigma}^0_{LR} \nonumber \\
& = &\frac{g_s^4}{27\pi\hat{s}} ( 1 +\frac{2m_b^2}{s} ) \nonumber \\
\hat{\sigma}^0_{LL}& = &\hat{\sigma^0}_{RR} = 0
\end{eqnarray}
At the tree level, the difference of the cross sections 
in the forward and backward directions is zero in each helicity state 
of $q\bar{q}$ because parity is conserved in QCD.

The one loop weak corrections to the $q\bar{q}$ cross section is 
\begin{eqnarray} 
\hat{\sigma}^1_{RL}& = &\hat{\sigma}^1_{LR} \nonumber \\
                   & = &\frac{g_s^4}{27\pi\hat{s}} 
                       [Re(A)( 2 +\frac{4m_b^2}{\hat{s}} )
                       +Re(C)(m_b)(-1 +\frac{4m_b^2}{\hat{s}})]
\end{eqnarray}
And the difference of $q\bar{q}$ cross sections 
in the forward and backward directions is 
\begin{eqnarray} 
\Delta\hat{\sigma}_{RL}& = &-\Delta\hat{\sigma}_{LR} \nonumber \\
                       & = &-\frac{g_s^4}{18\pi\hat{s}} \beta^2 [Re(B)]
\end{eqnarray}
where $\beta=\sqrt{1-4m^2_b/\hat s}$ and $A$, $B$ and $C$ 
are form factors defined in Eq.~\ref{eq:FF1}. 
A nonzero $\Delta\hat{\sigma}_{RL}$ is of course 
a signature of parity violation.

In polarized $pp$ and $p\bar{p}$ collisions, 
the differential cross section for the production of $b\bar{b}$ 
takes the form 
\begin{eqnarray} 
\frac{d\sigma}{dx_1 dx_2}
& = & \hat{\sigma}_{RL}
     [q_R(x_1,\hat{s})\bar{q}_L(x_2,\hat{s}) +
q_R(x_2,\hat{s})\bar{q}_L(x_1,\hat{s}] 
     \nonumber \\
&  +& \hat{\sigma}_{LR}
     [q_L(x_1,\hat{s})\bar{q}_R(x_2,\hat{s}) +
q_L(x_2,\hat{s})\bar{q}_R(x_1,\hat{s}] 
     \nonumber \\
&  +& \hat{\sigma}_{RR}
     [q_R(x_1,\hat{s})\bar{q}_R(x_2,\hat{s}) +
q_R(x_2,\hat{s})\bar{q}_R(x_1,\hat{s}] 
     \nonumber \\
&  +& \hat{\sigma}_{LL}
     [q_L(x_1,\hat{s})\bar{q}_L(x_2,\hat{s}) +
q_L(x_2,\hat{s})\bar{q}_L(x_1,\hat{s}] 
\end{eqnarray}
where $x_1$ and $x_2$ are momentum fractions 
of the partons in the initial beam.

Parity violation can be studied at hadron colliders by using polarized
initial beams, e.g. $p_R p_L$ or $p_R \bar{p}_L$ etc. 
Let us consider two hadron colliders with 
polarized beams (a) $p_R p_L$ and (b) $p_R \bar{p}_L$.
With polarized $pp$ or $p\bar{p}$ beams, 
we can define $q_i^+(x,\hat{s})$ [$q_i^-(x,\hat{s})$] 
to be the probability density \cite{Keller} for a quark of flavor $i$ 
and momentum fraction $x$  with a helicity of the same [opposite] sign 
as the helicity of the proton. 
The differential cross section, $d\sigma_{RL}$ 
in polarized $p_R p_L$ collisions become 
\begin{eqnarray} 
\frac{d\sigma_{RL}}{dx_1 dx_2}
& = & \frac{d\sigma_{LR}}{dx_1 dx_2} \nonumber \\ 
& = & \hat{\sigma}_{RL}
      [q^+(x_1,\hat{s})\bar{q}^+(x_2,\hat{s}) +
q^-(x_2,\hat{s})\bar{q}^-(x_1,\hat{s})
       \nonumber \\
&   & +q^-(x_1,\hat{s})\bar{q}^-(x_2,\hat{s}) +
q^+(x_2,\hat{s})\bar{q}^+(x_1,\hat{s})] 
      \nonumber \\
\sigma_{RL} & = &
\int^{1}_{x_1^{min}} dx_1 \int^{1}_{x_2^{min}} dx_2 
\frac{d\sigma_{RL}}{dx_1 dx_2}
\label{eq:sigma}
\end{eqnarray}
For $gg \to b\bar{b}$, we need to sum over all helicity states: 
$g_R g_L$, $g_L g_R$ $g_R g_R$ and $g_L g_L$.

The difference of $\sigma$ in the forward ($0 < \theta < \pi$) 
and the backward ($-\pi < \theta < 0$) directions 
at the $b\bar{b}$ CM frame 
in polarized $p_{\lambda_1}p_{\lambda_2}$ collisions are
\begin{eqnarray} 
\frac{\delta\sigma_{RL}}{dx_1 dx_2}
& = &-\frac{\delta\sigma_{LR}}{dx_1 dx_2} \nonumber \\ 
& = & \Delta\hat{\sigma}_{RL}
      [q^+(x_1,\hat{s})\bar{q}^+(x_2,\hat{s}) -
q^-(x_2,\hat{s})\bar{q}^-(x_1,\hat{s})  
       \nonumber \\
&   & -q^-(x_1,\hat{s})\bar{q}^-(x_2,\hat{s}) +
q^+(x_2,\hat{s})\bar{q}^+(x_1,\hat{s})] 
       \nonumber \\
\Delta\sigma_{RL} & = &
\int^{1}_{x_1^{min}} dx_1 \int^{1}_{x_2^{min}} dx_2 
\frac{\delta\sigma_{RL}}{dx_1 dx_2}
\label{eq:dsigma}
\end{eqnarray}
At high energy, parity violation from chromo-anapole form factor 
is expected to increase with $M_{b\bar{b}}$. 
It is therefore useful to examine 
the differential forward-backward asymmetry given in Eq.~\ref{eq:deltaA}.
The numerator has contribution only from $q\bar{q} \to b\bar{b}$, 
while the denominator has contributions from both $q\bar{q} \to b\bar{b}$ 
and $gg \to b\bar{b}$. 
Gluon fusion produces a large number of $b\bar{b}$ pairs,
which tend to contribute significantly to the denominator of  
Eq.~\ref{eq:deltaA} thus reducing the signal 
for the parity violating asymmetry
from $q\bar{q}$ in $pp$ collisions. 
In $p\bar{p}$ collisions, the antiquark density is greatly enhanced,
therefore, parity violation signals from $q\bar{q} \to b\bar{b}$ 
is much larger in $p\bar{p}$ than in $pp$ collisions at the same energy.
This differential asymmetry is presented in Figures 2 and 3 for polarized 
$p_R p_L$ and $p_R \bar{p}_L$ collisions at several values of energy.
The parity violation signal peaks at the $M_{b\bar{b}} = 2 m_t$ 
in the differential of forward-backward asymmetry versus $M_{b\bar{b}}$.
Also shown is the same asymmetry in $q\bar{q} \to Z \to b\bar{b}$; 
it dominates if $M_{b\bar{b}}$ is close to $M_Z$, but becomes negligible 
for $| M_{b\bar{b}} -M_Z | > 10$ GeV.

The integrated forward-backward asymmetry is defined as 
\begin{eqnarray}
{\cal A}  & \equiv & \frac{N_{F} -N_{B} }{ N_{F} +N_{B} }
                    =\frac{\sigma_{F}-\sigma_{B} }{\sigma_{F}+\sigma_{B}}
                     \nonumber \\
N & = & \cal{L} \sigma
\label{eq:AFB}
\end{eqnarray}
where $N_F$ and $N_B$ are the number of $b\bar{b}$ pairs 
in the forward and the backward directions; 
and $\cal{L}$ is the integrated luminosity.
The statistical uncertainty ($\Delta{\cal A}$)
and the statistical significance ($N_S$) of the integrated asymmetry are
\begin{eqnarray}
\Delta{\cal A} & \simeq & \frac{1}{ \sqrt{N_{F}+N_{B}} } \label{DeltaA} \\
N_S & \equiv & {\cal A}/\Delta{\cal A}.
\end{eqnarray}

The difference ($\Delta\sigma$)  and the total ($\sigma$)
of the cross sections $\sigma_F$ and $\sigma_B$
with one loop weak corrections,
as well as the integrated asymmetry (${\cal A}$)
and its statistical significance ($N_S$) are presented in Table 1 
for an integrated luminosity of 10 fb$^{-1}$ and 100 fb$^{-1}$.
For $\sqrt{s} \ge 1$ TeV, this  asymmetry might be visible
with $\cal{L} =$ 100 fb$^{-1}$ and a cut on $M_{b\bar{b}}$.
We find that requiring a higher $M_{b\bar{b}}$ 
can efficiently enhance the asymmetry from weak corrections 
in $q\bar{q} \to b\bar{b}$. 
A high $M_{b\bar{b}}$ cut 
not only reduces the cross section from gluon fusion 
it also reduces the asymmetry from $q\bar{q} \to Z \to b\bar{b}$ as well.

We close with two brief remarks.

1) In our calculations 
we have retained only the parity violating effects 
originating from $q\bar{q} \rightarrow b\bar{b}$. 
In principle, gluon fusion, i.e. $gg \rightarrow b\bar{b}$, 
can also contribute to parity violation 
due to electroweak corrections \cite{WIP}. 
However, dimensional reasoning indicates that 
such contributions are probably small at lower energy, 
at least for $\sqrt{\hat{s}} \lsim 2m_t$.
In any case, it is very unlikely that 
the contribution to the forward backward asymmetry from such a source 
will cancel away the $q\bar{q}$ contribution 
for all values of $M_{b\bar{b}}$. 

2) It would clearly be extremely interesting 
to consider these parity violating effects 
in other extensions of the SM; in particular in the MSSM. 
As we noted before, in MSSM, there will be
additional loop contributions not contained in a THDM that could
enhance the asymmetry \cite{WIP}.

%-----------------------------------------------------------------------
%   4. Conclusions
%-----------------------------------------------------------------------
\section{Conclusions}

To summarize, at a hadron collider with polarized $pp$ and $\sqrt{s} = 500$
GeV (e.g.\ RHIC), parity violation in $b\bar{b}$ production 
will be dominated by $q\bar{q} \to Z \to b\bar{b}$.  
In polarized $p_R \bar{p}_L$ collisions with $\sqrt{s} \ge 1000$ GeV 
and enough integrated luminosity, 
it might be possible to observe parity violation signals 
from SM weak corrections in $b\bar{b}$ production.
In the THDM with Model II Yukawa interactions, the chromo-anapole form factor 
of the bottom quark generated from leading weak corrections 
is enhanced for $\tan\beta$ close to one. 
It is enhanced by the factor $\cot^2\beta$ for $\tan\beta$ less than one. 
Therefore, the signal of parity violation in $b\bar{b}$ production 
is greatly enhanced for $\tan\beta \lsim 1$ 
and $M_{H^+}$ less than about 300 GeV.\footnote{
In the same model, parity violation in $t\bar{t}$ production \cite{Top}, 
is greatly enhanced for $\tan\beta \gsim m_t/m_b$.
In Ref. \cite{Top}, parity violation in $t\bar{t}$ production 
was studied in unpolarized $p\bar{p}$ collisions.}
The study of parity violation in $b\bar{b}$ production 
might provide a good opportunity to study new interactions
between the third generation quarks and (charged) spin-0 
as well as spin-1 bosons.

%-----------------------------------------------------------------------
%   THE ACKNOWLEDGEMENTS 
%-----------------------------------------------------------------------

\section*{Acknowledgements}

We are grateful to Stephane Keller and Jeff Owens for providing 
polarized parton distribution functions and for beneficial discussions.
C.K. and D.A. would like to thank Bill Bardeen 
and the Theoretical Physics Department of Fermilab for hospitality 
where part of this research was completed.
A.S. would like to thank Mike Tannenbaum for interesting 
discussions.
This research was supported in part by the the U.S. Department of Energy 
Grant Nos. 
DE-FG05-87ER40319 (Rochester), 
DE-FG02-95ER40896 (Wisconsin), 
DE-AC05-84ER40150 (CEBAF), 
and DE-AC-76CH00016 (BNL), 
and in part by the University of Wisconsin Research Committee 
with funds granted by the Wisconsin Alumni Research Foundation.

%-----------------------------------------------------------------------
%   THE BIBLIOGRAPHY
%-----------------------------------------------------------------------
\newpage
%

%-----------------------------------------------------------------------
%   TABLE CAPTIONS
%-----------------------------------------------------------------------
\newpage
\section*{Tables}

\bigskip

Table 1. 
The difference ($\Delta \sigma$) and the total ($\sigma$) 
of cross sections $\sigma_F$ and $\sigma_B$, 
and the asymmetry of $p_R \bar{p}_L \to b\bar{b} +X$ 
as defined in Eq.~\ref{eq:AFB} generated by weak corrections 
in (a) the SM and (b) the THDM 
with $M_{H^+} = 200$ GeV and $\tan\beta = 1$, 
for $\sqrt{s} = 500$ GeV with $M_{b\bar{b}} > $ 100 GeV, 
$\sqrt{s} = 1$ TeV with $M_{b\bar{b}} > $ 200 GeV and 
$\sqrt{s} = 2$ TeV with $M_{b\bar{b}} > $ 300 GeV.
Also shown is the statistical significance ($N_S$)
for ${\cal L} =$ 10 fb$^{-1}$ and 100 fb$^{-1}$.

\bigskip

\begin{center}
\begin{tabular}{ccccccc}
\hline
$\sqrt{s}$ & $\Delta\sigma$ & $\sigma(q\bar{q})$ & $\sigma(gg)$ 
& ${\cal A}$  & $N_S$ & $N_S$ \\
(GeV) & (pb) & (pb) & (pb) & ($\%$)  & (10 fb$^{-1}$) & (100 fb$^{-1}$) \\
\hline
(a) SM \\
500  & -0.130 & 124  & 81.2 & -0.063 & 0.91 & 2.9 \\
1000 & -0.081 & 22.9 & 17.2 & -0.20  & 1.3  & 4.1 \\
2000 & -0.122 & 16.6 & 33.3 & -0.25  & 1.7  & 5.5 \\
(b) THDM \\
500  & -0.156 & 124  & 81.2 & -0.076 & 1.1 & 3.4 \\
1000 & -0.106 & 22.9 & 17.2 & -0.27  & 1.7 & 5.3 \\
2000 & -0.163 & 16.7 & 33.3 & -0.33  & 2.3 & 7.3 \\
\hline
\end{tabular}
\end{center}
%

%-----------------------------------------------------------------------
%   FIGURE CAPTIONS
%-----------------------------------------------------------------------
\newpage
\section*{Figures}

\bigskip

\noindent Fig. 1. 
The real part of the chromo-anapole form factor $B(\hat{s})$  
as a function of $M_{b\bar{b}}$
from (a) the diagrams with $G^+$ (dash), the $W^+$ (dot-dash), 
the $Z$ (dot) and the SM Total; 
and the THDM Total for $\tan\beta =$ 0.5, 1, 3, 10 and 35, 
with (b) $m_{H^+} = 200$ GeV and (c) $m_{H^+} = 400$ GeV.   
Note that $\hat{s}=M_{b\bar{b}}$. 

\bigskip

\noindent Fig. 2. 
The differential forward-backward asymmetry ($\delta A$) 
defined in Eq.~\ref{eq:deltaA} 
versus $M_{b\bar{b}}$, in polarized $p_R p_L$ collisions for 
(a) $\sqrt{s} =  500$ GeV
(b) $\sqrt{s} = 1000$ GeV and (c) $\sqrt{s} = 2000$ GeV. 
This asymmetry has been evaluated using the (total) chromo-anapole form factor 
from leading weak corrections in the SM 
as well as in the THDM for $m_{H^+} = 200$ GeV 
and $\tan\beta =$ 0.5, 1, and 35.

\bigskip

\noindent Fig. 3. 
The differential forward-backward asymmetry ($\delta A$) 
defined in Eq.~\ref{eq:deltaA}
versus $M_{b\bar{b}}$, in polarized $p_R \bar{p}_L$ collisions for 
(a) $\sqrt{s} =  500$ GeV
(b) $\sqrt{s} = 1000$ GeV and (c) $\sqrt{s} = 2000$ GeV. 
This asymmetry has been evaluated 
using the (total) chromo-anapole form factor 
from leading weak corrections in the SM 
as well as in the THDM for $m_{H^+} = 200$ GeV 
and $\tan\beta =$ 0.5, 1, and 35. 
%

%=======================================================================
%   END DOCUMENT
%=======================================================================
\end{document}